\documentclass[12pt,preprint]{aastex}
\newcommand\lsim{\lower0.5ex\hbox{$\; \buildrel < \over \sim \;$}}
\newcommand\gsim{\lower0.5ex\hbox{$\; \buildrel > \over \sim \;$}}

\newcommand\ms{\mbox{$M_{\odot}$}}

\newcommand\ergs{\mbox{${\rm erg\,s}^{-1}$}}
\newcommand\my{\mbox{$\ms{\rm yr}^{-1}$}}
\newcommand\mbh{\mbox{$M_{\rm BH}$}}
\shorttitle{Evolution of IMBH X-Ray Binaries} \shortauthors{Li}

\begin{document}

\title{Evolution of Intermediate-Mass Black Hole X-Ray
Binaries}

\author{Xiang-Dong Li}

\affil{Department of Astronomy, Nanjing University, Nanjing
210093, P. R. China}
\email{lixd@nju.edu.cn}

\begin{abstract}
The majority of the ultraluminous X-ray sources (ULXs) in external
galaxies are believed to be accreting black holes in binary
systems; some of the black holes could be as massive as $\sim
100-1000\,\ms$. We have performed evolution calculations for
intermediate-mass black hole X-ray binaries, assuming they are
formed in dense star clusters via tidal capture. The results are
compared with those for stellar-mass black holes X-ray binaries.
We find that these two types of black holes may have similar
companion stars and binary orbits if observed as ULXs. However,
intermediate-mass black holes seem to be favored in explaining the
most luminous ULXs. We also discuss the possibilities of transient
behavior and beamed emission in the evolution of these binary
systems.

\end{abstract}


\section{Introduction}

Ultraluminous X-ray sources (ULXs) are point-like extra-nuclear
X-ray sources found in nearby galaxies, with (isotropic) X-ray
luminosities in excess of $10^{39}$ ergs$^{-1}$, roughly
corresponding to the Eddington limit accretion luminosity for a 10
$\ms$ star \citep{f89}. The majority of ULXs are believed to be
black holes accreting from their binary companion stars
\citep[see, e.g., reviews by][]{fw04,cm04}, or in few cases,
perhaps from the fallback material originating from supernovae
that have produced these sources \citep{li03}. Many works have
shown that stellar-mass black holes (SMBHs) can account for the
properties of  most ULXs, but for the most luminous ones (with
X-ray luminosities $L_{\rm X}\gsim 10^{40}\,\ergs$), assumptions
of either truly super-Eddington X-ray emission \citep{b02}, or
anisotropic (beamed) emission \citep{k01,kfm02} are required.
Alternatively, the latter sources could be intermediate-mass black
holes \citep[IMBHs;][]{cm99}. This suggestion seems to be
supported by their relatively low disk color temperatures inferred
from X-ray spectral analysis \citep[e.g.][]{mfm04}.

It is interesting and important to discriminate between these two
kinds of black holes observationally. \citet{k04} suggested that
IMBHs in X-ray binaries are more likely to be transient due to the
thermal-viscous instability \citep{kkb96}. The reason is that,
both the high masses and wide orbits of IMBHs lead to a larger
accretion disk and hence lower temperature in the disk compared to
the SMBH case. In this Letter, with different assumptions on IMBH
binary formation from \citet{k04}, we have calculated the
evolution of IMBH X-ray binaries with a massive donor star. We
describe the initial parameters for the binary evolution and the
calculated results in section 2 and 3 respectively. In section 4
we discuss the relation between IMBHs, SMBHs, and ULXs. Our
conclusions are summarized in section 5.

\section{Assumptions on the formation of IMBH X-ray binaries}

We consider IMBHs as ULXs associated with star formation regions
in spiral or irregular galaxies. In this situation, an IMBH could
be formed through runaway collision in dense young star clusters
within $\sim 3$ Myr \citep{pz04}. It may acquire a binary
companion via exchange encounters and/or tidal capture in the host
cluster. The latter process was recently investigated by
\citet{hpza04}. Some of their results were adopted in our
evolution calculations.

The BH masses were taken to be $\sim 100-1000\,\ms$ - more massive
BHs were not considered here. The reasons are as follows. First,
both spectral analysis \citep[e.g.][]{m03,w04,ds04} and numerical
calculations \citep{pz04} have revealed IMBH masses within this
range. Second, successful circularization of binaries formed by
tidal capture is only possible provided that the BH masses are
less than a few thousand $\ms$ \citep{hpza04}.

The masses of the donor stars were assumed to be $\sim 5-20\,\ms$.
These massive stars sink to the center of the clusters together
with the BH, and are more likely to be captured by a BH because of
their large cross section and close spatial distribution around
the BH \citep{k04}.

After in-spiral and circulation, the binary separation can be
derived from orbital angular momentum conservation to be
\citep{hpza04},
\begin{equation}
a\sim (4-5)(\mbh/M)^{1/3}R,
\end{equation}
where $\mbh$ is the BH mass, $M$ and $R$ the mass and radius of
the companion star, respectively. The above equation implies that
the stellar radius is roughly half of its Roche-lobe radius, i.e.,
the orbital periods of the incipient binaries are around $2-4$
days.

\section{Results of binary evolution calculations}

We have followed the evolution of the binary systems containing an
IMBH and a massive donor star for the initial parameters given in
last section, using an updated version of the evolution code
developed by \citet{e71}. The opacities in the code are from
\citet{ri92}, and from \citet{af94} for temperatures below
$10^{3.8}$ K. For the donor star we assumed a solar chemical
composition ($X=0.7$, $Y=0.28$, $Z=0.02$) and a mixing length
parameter $\alpha=2$. To follow the details of mass transfer
process, we included losses of orbital angular momentum due to
mass loss and gravitational wave radiation. We limited the mass
accretion rate of the black hole to its Eddington limit rate, and
let the excess mass be lost from the system with the specific
orbital angular momentum of the black hole. We also assumed that
the companion stars are on zero-age main-sequence when they have
been captured and settled in a circular orbit. This means that the
time for {\em circular} IMBH binary formation is much less than
the stellar main-sequence lifetime. This may not be true for the
companion stars more massive than $\sim 15\,\ms$, since the
formation history of IMBH binaries could be as long as $\sim 10^7$
yrs \citep{pz04}. So our results for stars of $M\gsim 15\,\ms$
should be regarded as the most optimistic cases.

Figure 1 shows two examples of mass transfer sequences for a
binary containing a $1000\,\ms$ BH with a 5 and 15 $\ms$ donor
star, respectively [changing the BH masses (say, to $100\,\ms$)
does not alter the results considerably]. In the figure the mass
transfer rates have been converted into X-ray luminosities to be
compared with observations. The X-ray luminosities were calculated
according to the slim disk model by \citet{o02}, in which photon
trapping effect was included\footnote{We fit the numerical results
in (more realistic) model B in \citet{o02} by the formula $L_{\rm
X}/L_{\rm E}\simeq 0.1\dot{m}/(1+\dot{m}^{1/2}/3)$, where $L_{\rm
E}$ is the Eddington luminosity, and $\dot{m}=\dot{M}c^2/L_{\rm
E}$.}. The solid and dashed curves correspond respectively to
stable and unstable mass transfer in the accretion disk, according
to the criterion given in \citet{d99}.

Since the initial binary orbit is too wide for the companion star
to fill its Roche lobe, the mass transfer through Roche-lobe
overflow begins until the star evolves and expands after a time
labelled below the time-axis in the figure. The X-ray luminosities
are generally around $10^{40}\,\ergs$, comparable with those of
the most luminous ULXs. However, the stable X-ray lifetime is
generally $\sim 10^6$ yrs, much shorter than the main-sequence
lifetime of the donor stars. Transient behavior usually occurs
when the orbits become sufficiently wide. The absence of extreme
long-term variability in ULXs suggests that they are likely to be
persistent X-ray sources \citep{r04}. If it is correct, IMBHs can
be observed as ULXs at the former part of their evolutions [at
later time the mass transfer rates can also be very high when the
donor star is on (super)giant branch, but with very short
duration, see \citet{rpp04}]. There is a dip/gap in the mass
transfer rates when the core hydrogen in the star exhausts and the
star deviates from thermal equilibrium. Similar features have
already been found in the calculations by \citet{k04},
\citet{prh03}, and \citet{rpp04}.

For comparison, we show in Fig.\,2 the mass transfer sequences for
typical SMBH X-ray binaries. The BH mass was assumed to be
$10\,\ms$. The mass of the donor star was also taken to be 5 and
15 $\ms$, respectively. In each case we considered case A and B
mass transfer. It can be seen that there is sufficiently long time
($\sim 10^7-10^8$ yrs) for these binaries to appear as persistent
ULXs with $L_{\rm X}\sim 10^{39}\,\ergs$, but the X-ray lifetime
with $L_{\rm X}\gsim 10^{40}\,\ergs$ is around a few $10^4-10^6$
yrs, comparable with and even shorter than that for IMBHs.

Figure 3 summarizes the evolutions of massive IMBH X-ray binaries.
Various symbols have been used to describe the states of the
binaries: stars, rectangles and triangles represent $L_{\rm
X}\gsim 10^{40}\,\ergs$, $10^{39}\,\ergs<L_{\rm
X}<10^{40}\,\ergs$, and $L_{\rm X}\lsim 10^{39}\,\ergs$, filled
and open symbols indicate stable and unstable mass transfer state,
respectively. The figure demonstrates that IMBH binaries, if
observed as persistent ULXs, have relatively massive companions in
narrow orbits. This feature is quite similar to those for SMBH
X-ray binaries \citep{prh03, rpp04}

\section{Discussion}

The purpose of this work is to examine whether the hypothesized
IMBHs can reproduce the observed properties of (some of) the ULXs.
There have been several works on this subject in the literature.
In the binary evolution calculations we have used the results in
\citet{hpza04} on tidal capture to determine the initial
conditions for the incipient IMBH binaries. But we allowed stellar
expansion before Roche-lobe overflow, and got the following
results different from theirs: (1) the donor stars have already
evolved off main-sequence when mass transfer begins; (2) the X-ray
lifetime for IMBHs as ULXs is around $10^6$ yr, much shorter than
the main-sequence lifetime of the donor stars. These can present
constraints on the donor stars of IMBH-ULXs. For example, the age
of the star cluster MGG-11 in the irregular galaxy M82 is $7-12$
Myr. If the ULX M82 X-7 in this cluster \citep{mt99} is powered by
an IMBH of mass $\sim 1000\,\ms$, a $10-15\,\ms$ donor star is
required to account for the luminosity and age of the ULX. In a
more recent paper, \citet{pzdm04} performed detailed binary
evolution calculations of $2-15\,\ms$ stars which transfer mass to
a $100-2000\,\ms$ BH. These authors initialized the binary systems
with an (arbitrarily) selected orbital period at birth (ranging
from less than a day to a few hundred days). Their calculated mass
transfer rates and time with initial orbital period of the order
of a day are in general agreement with ours.

Systematic studies on evolution of SMBH X-ray binaries have been
carried by \citet{prh03} and \citet{rpp04}. The results in Fig.~2
can be regarded as individual cases of their population
synthesized results. We notice that different parameterization for
estimating the X-ray luminosities has been adopted here and in
their works. \citet{rpp04} assumed an energy conversion
coefficient depending on the BH spin and let $L_{\rm X}$ be as
high as 10 times of the Eddington luminosity, as suggested by
\citet{b02}. We have used the results of \citet{o02}, who
calculated the emission from the supercritical accretion disk,
considering the photon trapping effect. This difference causes the
mass transfer rate to be about 10 times higher in our work (i.e.,
$\dot{M}\sim 10^{-5}\,\my$) than in \citet{rpp04}, for $L_{\rm X}$
to reach $10^{40}\,\ergs$, for a $10\,\ms$ BH. The duration of the
high-luminosity X-ray emission phase is correspondingly $\sim 10$
times shorter to be $\sim 10^6$ yrs for a massive donor star.

From evolution calculations we have shown that, as ULXs, both IMBH
and SMBH X-ray binaries actually have similar characteristics in
their companions stars and binary orbits, which make it difficult
to discriminate them observationally. However, if the donor star
is more massive than $\sim 10\,\ms$, the IMBH binaries are most
likely to reside deeply inside the host star cluster, while SMBH
binaries generally have been "kicked" outside \citep{z02}. For an
intermediate- or low-mass donor star, the IMBH binaries could be
hostless, since the preceding time before mass transfer could be
long enough for the cluster to be tidally dispersed, if it is
close to the galactic center \citep{p04}.

For the most luminous ULXs with $L_{\rm X}\gsim 10^{40}\,\ergs$, a
potential problem related to the SMBH model  is that, according to
\citet{o02}, the effective temperatures of supercritical accretion
disks with $\dot{M}\sim 10^{-5}\,\my$ are still as high as $\sim
1$ keV, inconsistent with spectral analyzed results for these
sources \citep[e.g.][]{mfm04}. It is not known how the temperature
profile is in the \citet{b02} disk model. On the other hand, IMBHs
have the advantage of low disk temperatures because of their high
masses.

We also find that transient behavior is a common feature in late
evolutionary stages in these two types of binaries. It may be
difficult to tell IMBHs from SMBHs by transient behavior, as
suggested by \citet{k04}. The differences between our work and
\citet{k04} result from the following facts. (1) We have adopted
the BH masses no more than $1000\,\ms$ in the calculations. With
these values, one can actually draw from Fig.~1 in \citet{k04} the
similar results as ours. (2) We assumed tidal capture as the main
formation channel of IMBH binaries, from which we got the initial
orbital periods of $\sim 2-4$ days; \citet{k04} favor orbital
periods in excess of $\sim 100$ days, but the duration of the mass
transfer episode would decrease to be very short (less than a few
$10^4$ yrs), making them difficult to be observed.

It should be noted that the calculated mass transfer rates are
long-term, averaged ones. It is unclear how to relate these
secular mass transfer rates to observable instantaneous X-ray
luminosities, and to disk instability influenced by (short-term)
mass transfer and X-ray irradiation. Moreover, the mechanisms and
criterion for the thermal-viscous instability in irradiated
accretion disk are not yet fully understood. It is premature to
predict the instability occurrence from only calculated mass
transfer rates.

Another interesting feature is that, if anisotropic (or beamed)
emission is associated with mass transfer rates comparable to the
Eddington rate, as suggested before \citep[e.g.][]{k01}, our
calculations indicate that IMBHs may also have anisotropic
emission, since the mass transfer rates can be sufficiently high
to satisfy the condition above. However, anisotropic X-ray
emission is not preferred in our opinion for persistent ULXs,
since the most important stabilizing factor for disk instability,
the efficiency of X-ray irradiation, will be greatly reduced if
the emission is beamed (usually in the direction perpendicular to
the disk plane), and the ULXs would become transient sources.

\section{Summary}
We have calculated the evolutionary sequences of IMBH X-ray
binaries formed through tidal capture in dense star clusters, and
compared the results with those of SMBH binaries. We found that
IMBHs seem to be capable of explaining the nature of most luminous
ULXs, and their companion stars and binary orbits could be similar
to those of SMBH-ULXs. We suggest that transient behavior and
beamed emission may be not enough to distinguish between IMBHs and
SMBHs.

\acknowledgements I would like to thank Ranold Webbink for helpful
discussion, and the referee, Philipp Podsiadlowski for clarifying
comments. This work was supported by NSFC through grant number
10025314 and MSTC through grant number NKBRSF G19990754.

\begin{figure}
\plotone{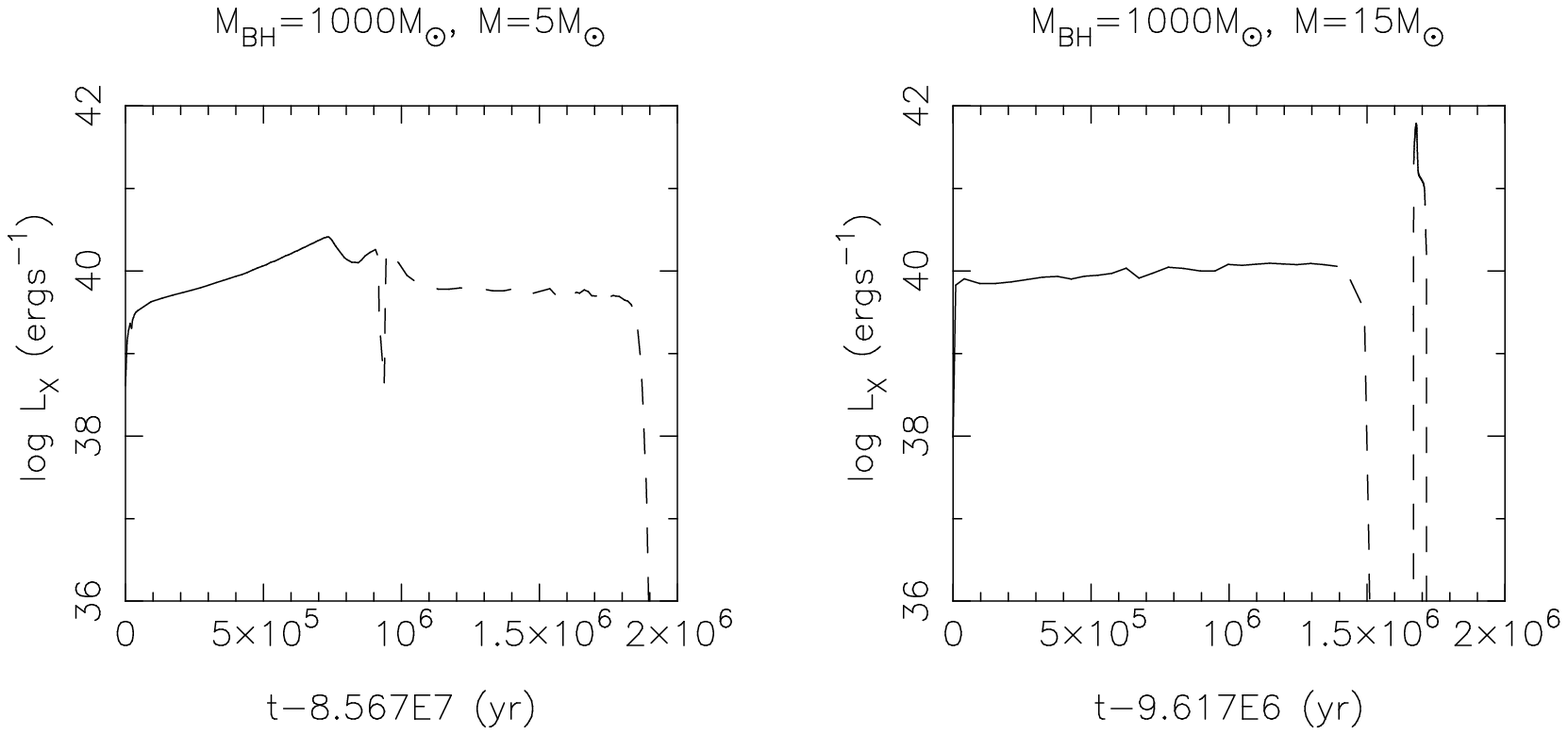} \caption{Potential X-ray luminosities as a
function of time for two IMBH X-ray binary evolution sequences.
The duration of the evolutionary phase before Roche-lobe overflow
has been subtracted from the real time. The solid and dashed
curves reflect thermal-viscously stable and unstable mass
transfer, respectively.
 \label{fig1}}
\end{figure}
\clearpage
\begin{figure}
\plotone{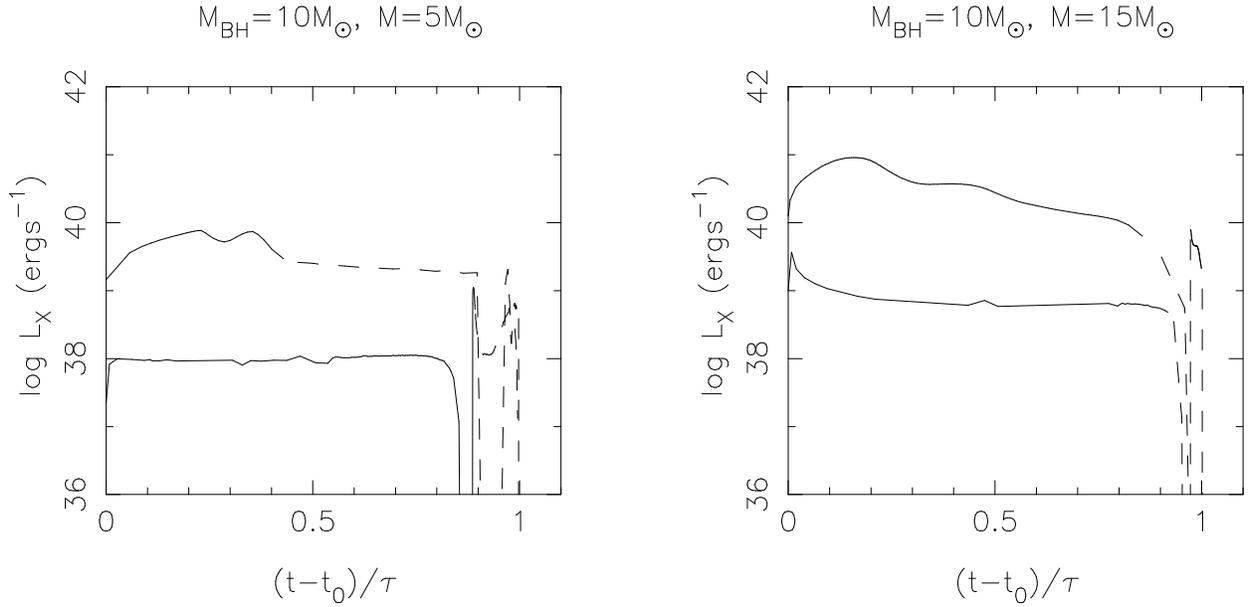} \caption{Potential X-ray luminosities as a
function of normalized time for four SMBH X-ray binary evolution
sequences. The solid and dashed curves reflect thermal-viscously
stable and unstable mass transfer, respectively. In the left panel
the parameters for the upper and lower curve are: the initial
binary orbital periods $P_{\rm orb,i}=3.43$, 0.87 d, the time at
the onset of mass transfer $t_0=1.0\times 10^8$, $3.8\times 10^7$
yr, and the duration of total mass transfer episode
$\tau=1.2\times 10^6$, $1.4\times 10^8$ yr, respectively. In the
right panel: $P_{\rm orb,i}=3.17$, 0.65 d; $t_0=1.1\times 10^7$,
$2.5\times 10^5$ yr; $\tau=4.4\times 10^4$, $3.1\times 10^7$ yr.
 \label{fig2}}
\end{figure}
\clearpage
\begin{figure}
\plotone{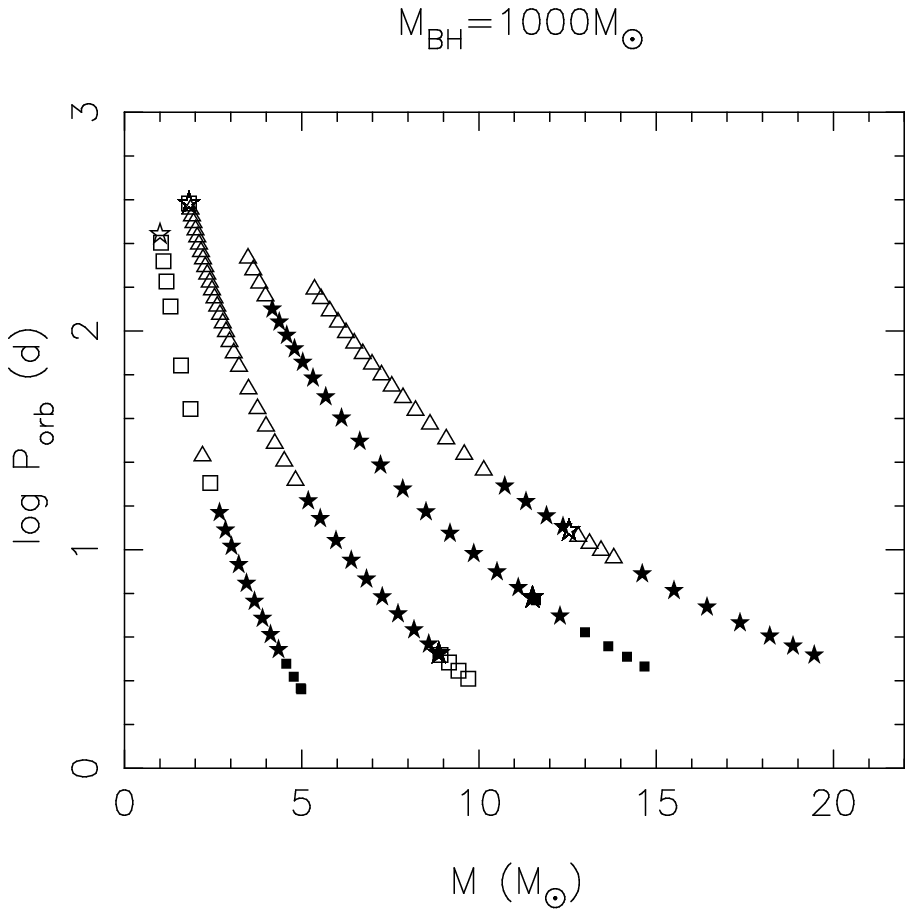} \caption{IMBH binary evolution sequences on the
donor mass - orbital period plane. See text for the meanings of
various symbols
 \label{fig3}}
\end{figure}


\begin{thebibliography}{}

\bibitem[Alexander \& Ferguson(1994)]{af94}Alexander, D. R. \&
Ferguson, J. W. 1994, ApJ, 437, 879
\bibitem[Begelman(2002)]{b02}Begelman, M. C. 2002, ApJ, 568, L97
\bibitem[Colbert \& Miller(2004)]{cm04}Colbert, E. J. M. \&
Miller, M. C. 2004, talk at the Tenth Marcel Grossmann Meeting on
General Relativity, Rio de Janeiro, July 20-26, 2003. Proceedings
edited by M. Novello, S. Perez-Bergliaffa and R. Ruffini, World
Scientific, Singapore, 2004 [astro-ph/0402077].
\bibitem[Colbert \& Mushotzky(1999)]{cm99}Colbert, E. J. M.
\& Mushotzky, R. F. 1999, ApJ, 519, 89
\bibitem[Di Stefano et al.(2004)]{ds04}Di Stefano, R., Primini, F.
A., Kong, A. K. H., \& Russo, T. 2004, preprint [astro-ph/0405238]
\bibitem[Dubus et al.(1999)]{d99}Dubus, G., Lasota, J.-P.,
Hameury, J.-M., \& Charles, P. 1999, MNRAS, 303, 139
\bibitem[Eggleton(1971)]{e71}Eggleton, P. P. 1971, MNRAS, 151, 351
\bibitem[Fabbiano(1989)]{f89}Fabbiano, G. 1989, ARA\&A, 27, 87
\bibitem[Fabbiano \& White(2004)]{fw04}Fabbiano, G. \& White, N.
E. 2004, to appear in "Compact Stellar X-ray Sources", eds. W. H.
G. Lewin and M. van der Klis (Cambridge Univ. Press, Cambridge)
[astro-ph/0307077]
\bibitem[Hopman, Portegies Zwart, \& Alexander(2004)]{hpza04}Hopman, C.
Portegies Zwart, S. F., \& Alexander, T. 2004, ApJ, 604, L101
\bibitem[Kalogera et al.(2004)]{k04}Kalogera, V., Henninger, M.,
Ivanova, N. \& King, A. R. 2004, ApJ, 603, L41
\bibitem[King et al.(2001)]{k01}King, A. R., Davies, M. B.,
Ward, M. J., Fabbiano, G., \& Elvis, M. 2001, ApJ, 552, L109
\bibitem[King, Kolb, \& Burderi(1996)]{kkb96}King, A. R., Kolb, U.,
\& Burderi, L. 1996 ApJ, 464, L127
\bibitem[K\"ording, Falcke, \& Markoff(2002)]{kfm02}K\"ording, E.,
Falcke, H., \& Markoff, S. 2002, A\&A, 382, L13
\bibitem[Li(2003)]{li03}Li, X.-D. 2003, ApJ, 596, L199
\bibitem[Matsumoto \& Tsuru(1999)]{mt99}Matsumoto, H. \& Isuru, T.
G. 1999, PASJ, 51, 321
\bibitem[Miller et al.(2003)]{m03}Miller, J. M., Fabbiano, G.,
Miller, M. C., \& Fabian, A. C. 2003, ApJ, 585, L37
\bibitem[Miller, Fabian, \& Miller(2004)]{mfm04}Miller, J. M.,
Fabian, A. C., \& Miller, M. C. 2004, preprint [astro-ph/0406656]
\bibitem[Ohsuga et al.(2002)]{o02}Ohsuga, K., Mineshige, S., Mori,
M., \& Umemura, M. 2002, ApJ, 574, 315
\bibitem[Podsiadlowski, Rappaport, \& Han(2003)]{prh03}Podsiadlowski, P.,
Rappaport, S., \& Han, Z. 2003, MNRAS, 341, 385
\bibitem[Portegies Zwart(2004)]{p04}Portegies Zwart, S. 2004,
to appear in "Joint Evolution of Black Holes and Galaxies" of the
Series in High Energy Physics, Cosmology and Gravitation, eds M.
Colpi, V.Gorini, F.Haardt and U.Moschella (IOP Publishing, Bristol
and Philadelphia) [astro-ph/0406550]
\bibitem[Portegies Zwart et al.(2004)]{pz04}Portegies Zwart, S.
F., Baumgardt, H., Hut, P., Makino, J., \& McMillan, L. W. 2004,
Nat, 428, 724
\bibitem[Portegies Zwart, Dewi, \& Maccarone(2004)]{pzdm04}Portegies Zwart, S.
F., Dewi, J., \& Maccarone, T. 2004, preprint [astro-ph/0408402]
\bibitem[Rappaport, Podsiadlowski, \& Pfahl(2004)]{rpp04}Rappaport, S.,
Podsiadlowski, P., \& Pfahl, E. 2004, preprint [astro-ph/0408032]
\bibitem[Roberts et al.(2004)]{r04}Roberts, T. P., Warwick, R. S., Ward, M.
J., \& Goad, M. R.  2004, MNRAS, 349, 1193
\bibitem[Rogers \& Iglesias(1992)]{ri92}Rogers, F. J. \& Iglesias,
C. A. 1992, ApJS, 79, 597
\bibitem[Wang et al.(2004)]{w04}Wang, Q. D., Yao, Y., Fukui, W., Zhang, S. N.,
\& Williams, R. 2004, ApJ, 609, 113
\bibitem[Zezas et al.(2002)]{z02}Zezas, A., Fabbiano, G., Rots, A.
H., \& Murray, S. S. 2002, ApJ, 577, 710

\end{thebibliography}
\end{document}